\documentclass{article}

\PassOptionsToPackage{numbers}{natbib}  
\usepackage[main, final]{neurips_2026}

\usepackage[utf8]{inputenc} 
\usepackage[T1]{fontenc}    
\usepackage{hyperref}       
\usepackage{url}            
\usepackage{booktabs}       
\usepackage{amsfonts}       
\usepackage{nicefrac}       
\usepackage{microtype}      
\usepackage{xcolor}         
\usepackage{graphicx} 
\usepackage{main-defs}
\usepackage{float}

\title{Agentic AI Detector Co-design and Optimization in Vertically Integrated Full Simulations}

\author{%
  Wonyong Chung \\
  Princeton University \\
  Princeton, NJ 08544, USA \\
  \texttt{wonyongc@princeton.edu} \\
  \And
  Qibin Liu \\
  SLAC National Accelerator Laboratory\\
  Menlo Park, CA 94025, USA \\
  \texttt{qibin@slac.stanford.edu} \\
  \AND
  Liangyu Wu \\
  Stanford University\\
  Stanford, CA 94305, USA \\
  \texttt{liangyu.wu@stanford.edu} \\
  \And
  Julia Gonski \\
  SLAC National Accelerator Laboratory\\
  Menlo Park, CA 94025, USA \\
  \texttt{jgonski@slac.stanford.edu} \\
}

\begin{document}

\maketitle

\begin{abstract}
We present the first implementation of AI agents into the design and optimization of detectors in high-energy physics experiments via a bi-level optimization framework that vertically integrates detector geometry, front-end digitization, and high-level reconstruction algorithm parameters in full simulations. 
Using the example of a dual-readout, segmented crystal EM calorimeter, a detector concept whose specification targets an EM resolution of $3\%/\sqrt{E}$~\cite{Lucchini:2020bac}, we investigate the capabilities and value propositions of AI agents in the identification and reduction of key detector parameters and in the nonlinear traversal of design space.
We find that a frontier reasoning model, provided with general detector and optimization knowledge through natural-language context in a fixed prompt, can effectively propose detector geometries, interpret the resulting performance, and proactively suggest relevant directions for further investigation and improvement.
Here, we demonstrate an agent-assisted workflow that selects a design through staged optimization of three criteria and reaches the best measured geometry region with fewer evaluations than exhaustive enumeration. This study establishes the foundation for a future demonstration of the first fully AI-designed detector for future scientific facilities.
\end{abstract}


\section{Introduction}

High-energy physics (HEP) research today is organized around a set of key physics drivers, identified through global community strategic planning exercises such as Snowmass~\cite{butler2023report}, P5~\cite{p5_2023}, and the European Strategy for Particle Physics~\cite{CERN-ESU-2025-002}.
These topics include precision measurements of the Higgs boson, broad discovery reach for beyond-the-Standard Model physics that can explain the nature of dark matter or the expansion of the universe, and the origin of neutrino mass. 
Successfully addressing each of these drivers requires a robust and broad landscape of future experimental infrastructure ranging from high-energy colliders to astrophysical observatories.
Each of these types of experiments requires dedicated R\&D in advanced detector technologies and designs.

Artificial intelligence and machine learning (AI/ML) have become primary workhorses for nearly every aspect of these research efforts~\cite{albertsson2019machinelearninghighenergy,aarrestad2026buildingainativeresearchecosystem}. 
The next generation of HEP experiments will rely on AI/ML in fundamentally novel ways, integrated into the full experimental design cycle, ranging from the use of real-time AI-based readout for on-detector signal processing in radiation-hard front-end electronics~\cite{Yoo_2024, Yilmaz_2025_ML4PS}, to pattern-scoring and decision-making processes in rare-event triggers, to overall detector geometry design parameters. 
AI/ML also enables new methods of achieving optimal detector design, leveraging varying approaches such as differentiable detector elements~\cite{Baydin02012021, mode_collaboration_website, Aehle_2025}, reinforcement learning~\cite{qasim2024physicsinstrumentdesignreinforcement}, and synthetic data representations~\cite{chung2025synthetictrainingrepresentationbridging}.  

The introduction of AI agents that can autonomously reason about and execute workflows has dramatically accelerated progress in software development and closely adjacent fields but their introduction into HEP workflows is still in its infancy.
Recent works have established the ability of AI agents across two wide swathes of the traditional training ground for PhD physicists in HEP: 1) established but unfinished theory calculations in a specific regime~\cite{schwartz2026resummationcparametersudakovshoulder}, and 2) standard analyses in collider physics~\cite{Moreno:2026mqk, birk2026scientifichumanagentreproductionpipeline}.
This work adds the example of phenomenological detector design.
Using the example of a dual-readout, segmented crystal EM calorimeter for future colliders, we demonstrate the first application of an agentic-AI driven workflow based on a bilevel optimization framework designed to perform simultaneous optimization of detector geometry and reconstruction parameters for a given set of physics targets. 
In this study the agent guided the geometry selection within the staged workflow. Still, physics-informed human input is required to ensure relevance of the agent's proposed tasks to the experimental context.

\section{Methods}

\begin{figure}[!htbp]
\centering
   \includegraphics[width=0.7\textwidth]{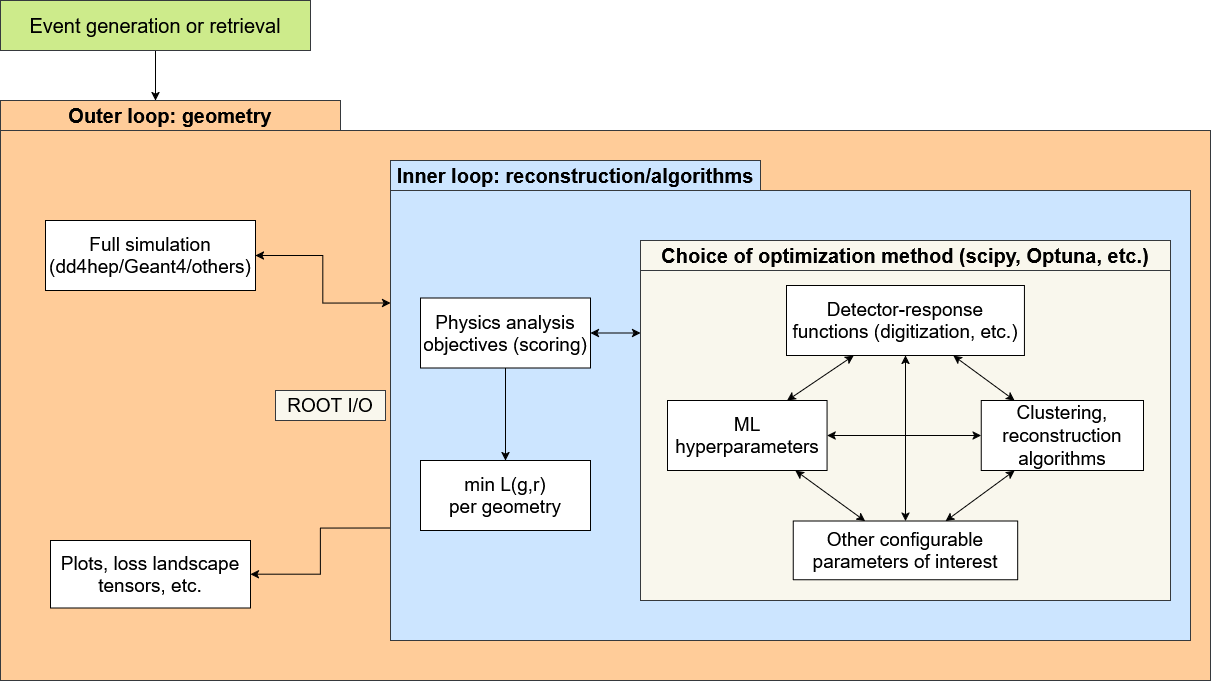}
    \caption{Bi-level detector optimization framework (VIBRATO), indicating geometry-based outer loop and reconstruction-based inner loop and their connection to user-defined optimization criteria.
    \label{fig:bilevel_arch}}
\end{figure}

This work debuts Vertically Integrated Bilevel geometry, Reconstruction, And TDAQ Optimizer (VIBRATO), a modular and extensible detector optimization framework~\cite{chung_bilevel_det_opt} that iterates \texttt{DD4HEP}~\cite{Frank_2014} full simulation runs with varying detector geometry parameters. 
It then evaluates simulation outputs against arbitrary downstream algorithm parameters such as front-end digitization or high-level reconstruction routines. Algorithm results are scored against a given physics- or statistics-objective and the final output is the loss landscape across the full geometry and algorithm parameter-space. Algorithms are designed to be plug-and-play with the full simulation outputs, with the only requirement that I/O be in ROOT~\cite{BrunRademakers1996ROOT} format. A menu of \texttt{scipy}~\cite{2020SciPy-NMeth} optimization methods are implemented for this study, with support for hyperparameter optimization for ML-methods in upcoming iterations. The plug-in interface is deliberately kept minimal to allow maximum flexibility.
Figure~\ref{fig:bilevel_arch} provides a diagram of the framework. 

Segmented dual-readout crystal electromagnetic (EM) calorimeters~\cite{RevModPhys.90.025002,hirosky2024dualreadoutcalorimetryhomogeneouscrystals,Eno:2025ltc} are a promising technology for delivering excellent jet energy resolution at a future collider.
Dual-readout calorimeters count the number of photons produced in the wake of a high-energy particle in both scintillation (S) and Cherenkov (C) channels, from which the EM fraction of hadronic showers, which suffers from a large intrinsic fluctuation due to asymptotic-freedom in hadronization, can be determined on an event-by-event basis. Calorimeters in colliders have historically only counted in the scintillation channel as a proxy for energy, and the addition of the Cherenkov channel, which counts the photons emitted in a distinctive wavefront as a particle travels faster than the speed of light in the medium, represents a genuinely new physical observable in detector hardware.
This study uses the segmented crystal ECAL of the IDEA detector concept~\cite{Lucchini:2020bac} and casts its design as a bilevel problem. The geometry parameters chosen to be varied are the nominal transverse-width parameter $W$, setting the transverse granularity, a projective radial offset $d$, and the front-segment crystal length $L_f$.

\begin{figure}[!htbp]
\centering
   \includegraphics[width=0.85\textwidth]{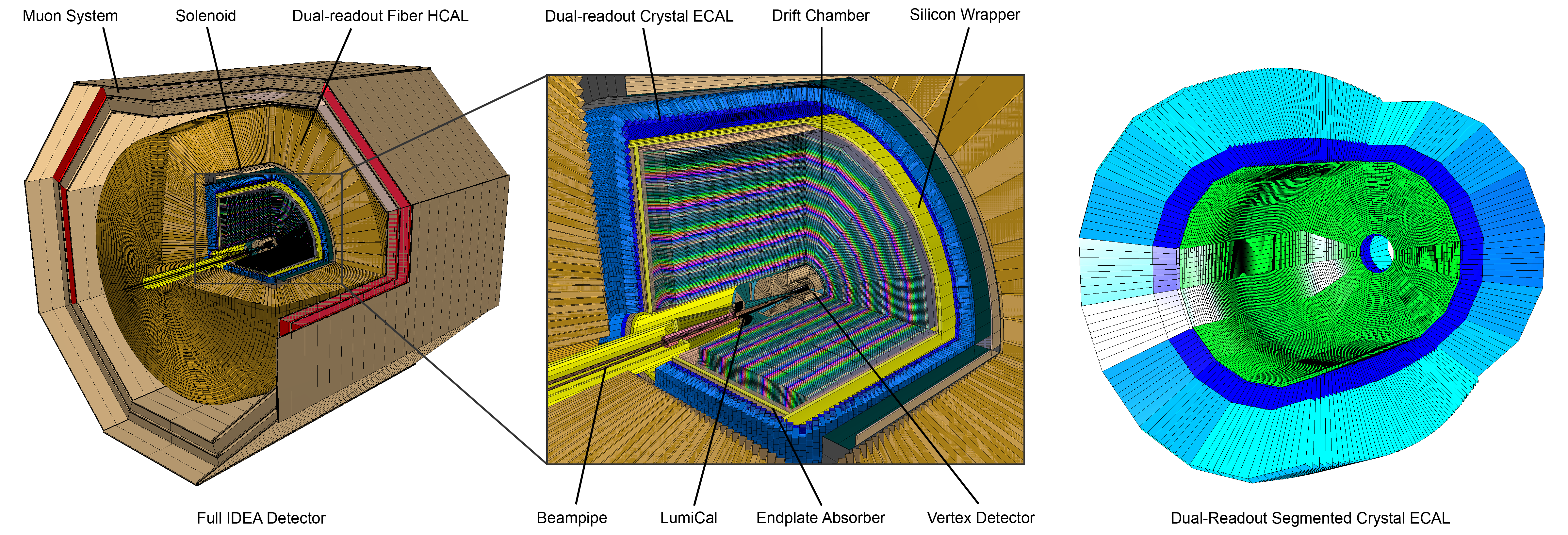}
    \caption{The IDEA detector concept with labeled subsystems (left, center). The Segmented Crystal EM Precision Calorimeter (SCEPCal) with exaggerated crystal sizes for visibility (right).
    \label{fig:idea_scepcal}}
\end{figure}

A \texttt{DD4HEP} full detector simulation running \texttt{Geant4}~\cite{AGOSTINELLI2003250} under the hood is performed for each geometry iteration. This domain-specific tool models particle-material interactions and records realistic energy-deposition information. A digitization algorithm then converts the simulated energy deposits into modelled detected-light counts and SiPM waveforms. Throughout this paper, ``S/C photon counts'' denote these modelled detected-light counts, parameterised from the energy deposits, rather than individually tracked optical photons. For each readout channel, the S/C photon counts are first extracted and aggregated with configurable granularity (per-crystal, per-layer, per-cluster, or per-event). Per-photon analog templates are then constructed analytically: the scintillation emission profile follows a two-component exponential decay model with a finite rise time, while the Cherenkov emission is modeled as a single fast exponential. Each emission profile is convolved via FFT with the SiPM single-photon response (SPR), which is parameterized as a double-exponential pulse shaped by a first-order RC high-pass filter representing AC coupling. The resulting per-photon templates are cached for reuse within each waveform-generation configuration.

Waveform synthesis proceeds by scaling the cached S/C templates by the respective photon counts, shifting them by a sampled signal arrival time $t_0$, then superimposes Poisson-distributed SiPM dark counts (each convolved with the SPR), additive Gaussian electronic noise, and a configurable DC baseline offset. The continuous analog waveform is sampled at a configurable rate determined by the sampling interval and readout time window. 
An electronics chain then applies ADC quantization by mapping the sampled signal to discrete codes within a specified dynamic range and bit depth. 

SciFi~\cite{liu2026scifisafelightweightuserfriendly} is a science-focused agentic-AI framework, a fully open-source workflow designed to offload and streamline common scientific workloads, including data processing, visualization, analysis through agentic loops, and interface them with scientific computing environments. It ensures safe execution and allows choice of the LLM backbone. Combined with the bilevel optimization framework above, which provides the domain-specific toolchain and incorporates detector-domain knowledge, this work demonstrates the first agent-driven optimization workflow for a realistic experimental detector design. 
The prompt design and point suggestion agent is a frontier reasoning model (\emph{Claude}, \texttt{claude-fable-5-1}) invoked afresh for every iteration: it receives only a fixed protocol prompt and the table of its own previous results, and it only proposes geometries and interprets the returned numbers. Simulation, the inner-loop optimisation, the bookkeeping of already simulated geometries and the digitization scan are executed around it by the toolchain. The search protocol (objective and its cost exponent, allowed values, statistics per geometry, epoch schedule and the rules for fixing and re-checking an axis) was fixed before the run.


The bilevel optimization is performed in a high-dimensional parameter space, where exhaustive enumeration is computationally intensive and correlations are present. Eleven parameters enter the workflow and are handled with different strategies. 
Five are design parameters that must be fixed before construction.
These include three geometry parameters, namely the nominal crystal width $W$ (the generator quantizes the azimuthal pitch, so the physical transverse dimensions at $W = 5$ cm are about $3.9 \times 5.5$ cm), the front-segment length $L_f$, and the projective offset $d$. 
The remaining two are digitization parameters, specifically the sampling rate $f_s$ and the ADC resolution $N$. 
The remaining six are nuisance parameters, with two for reconstruction: the clustering radius $R$ and the energy threshold $E_{thr}$, which the inner loop optimizes automatically for every geometry with a seeded differential-evolution search within physical bounds ($R$ at least one crystal pitch, $E_{thr} \le 5$ MeV; additional optimizer seeds serve as spread checks). 
The final four nuisance parameters are waveform-fit settings (the C and S fit bounds, the $t_0$ window and the $\chi^2$ quality cut) that are fixed for the optimization and checked afterwards at the final design point (Section~\ref{sec:results}).

The optimization target is multi-objective. The geometry objective is a cost-weighted energy signal-to-noise ratio, $\mathrm{SNR}/(C/100\,\mathrm{M\$})^{0.35}$, where SNR is the mean over all events of the clustered energy divided by $\sqrt{n_{\rm hits}}$ (events without a valid cluster count as zero and a clustering efficiency of at least 99\,\% is required) and $C$ is a modelled crystal-plus-readout cost: the total crystal volume at a price per volume with an assumed size premium based on the transverse dimensions and the combined front-plus-rear length, plus a per-channel readout cost. The exponent of 0.35 is chosen to reflect a particular preference in the performance–cost trade-off, without loss of generality.
The digitization objectives are the per-crystal mean absolute relative error on the fitted number of S photons from the waveform fit and the static ADC power, for which an assumed two-parameter model $P = N_{\rm ch}\,(P_{\rm FE} + \mathrm{FoM}\cdot 2^{N} f_s)$ is used, with a front-end floor $P_{\rm FE} = 10$ mW per channel and a figure of merit $\mathrm{FoM} = 30$ fJ per conversion step, summed over all channels.

\section{Results}
\label{sec:results}

The optimization was run under a fixed protocol with single 1 GeV electrons ($\theta = 60$--$120^\circ$, 2 T field, 1000 events per geometry) in the modelled SCEPCal readout.
The protocol has four proposal epochs of up to twelve geometries each and three separate agent trajectories were run with the same protocol.
For the present analysis the reported endpoint of a trajectory is the highest-scoring geometry of its last proposal epoch.
 Figure~\ref{fig:results_offset} shows one trajectory (the others are in Appendix~\ref{app:replicas}) on top of the landscape of all simulated geometries; the chosen reference point on the measured plateau is $(W, L_f, d) = (5, 10, 5)$ cm. 

In the geometry aspect, we find the projective offset $d$ is probed by the agent as a nuisance axis and the chosen value differs from run to run. 
The clustering parameters were optimized automatically by the inner loop at every proposed geometry. The allowed values, step sizes and event statistics were fixed by the protocol; which geometries to evaluate in each epoch was the agent's choice. 
The geometry objective is independent of the subsequent digitization and fit settings, so the digitization is optimized as a second stage, conditional on the selected geometry.
This decomposes the full 11-dimensional problem into tractable stages by design as a staged procedure (i.e. not a proof of joint optimality). The four fit settings were fixed for the optimization and checked at the final design point.

The 11-dimensional problem is thus handled in stages with a different strategy for each block: the agent first searches the three geometry parameters in the outer loop, temporarily concentrating on two of them while retaining checks of the offset.
Next, the inner loop optimizes the two clustering parameters for every geometry: the two digitization parameters are scanned at the chosen geometry with two objectives ($S_{\mathrm{photon}}$ error and ADC power), and the four fit settings are fixed and checked. 
Once the geometry is fixed, the sampling rate and ADC resolution are chosen as the point of lowest $S_{\mathrm{photon}}$ error under a static power budget of 5 kW, which exposes the trade-off with the third metric.
The reported design combines the geometry endpoint of the illustrated trajectory with the subsequent digitization selection: crystal width $W = 5$ cm, front-segment length $L_f = 10$ cm, projective offset $d = 10$ cm, sampling rate $f_s = 1$ GHz and ADC resolution $N = 10$ bit. The geometry gives a cost-weighted SNR of $0.347 \pm 0.002$.
The digitization gives a per-crystal mean absolute relative error on the number of S photons of 4.9\,\% 
at a static ADC power of 2.8 kW for 68,608 channels\footnote{The ADC power is an assumed static estimate, $P = N_{\rm ch}(10\,\mathrm{mW} + 30\,\mathrm{fJ}\cdot 2^{N} f_s)$: a 10 mW per-channel front-end floor plus a figure of merit of 30 fJ per conversion step. It should be read as a relative cost metric for design optimization; the absolute values require calibration against device-level implementations.}. The two other trajectories end at the same width and length, $W = 5$ cm and $L_f = 10$ cm, with $d = 15$ cm (Appendix~\ref{app:replicas}). 

Each trajectory used 48 geometry evaluations.
Given that at each configuration a relatively large amount of computation is required, the problem presented here is already costly to enumerate exhaustively. Nonetheless, expert physicist input is required to ensure the agent remains in technically and practically realizable design space.


\begin{figure}[!htbp]
    \centering
    \includegraphics[width=0.47\textwidth]{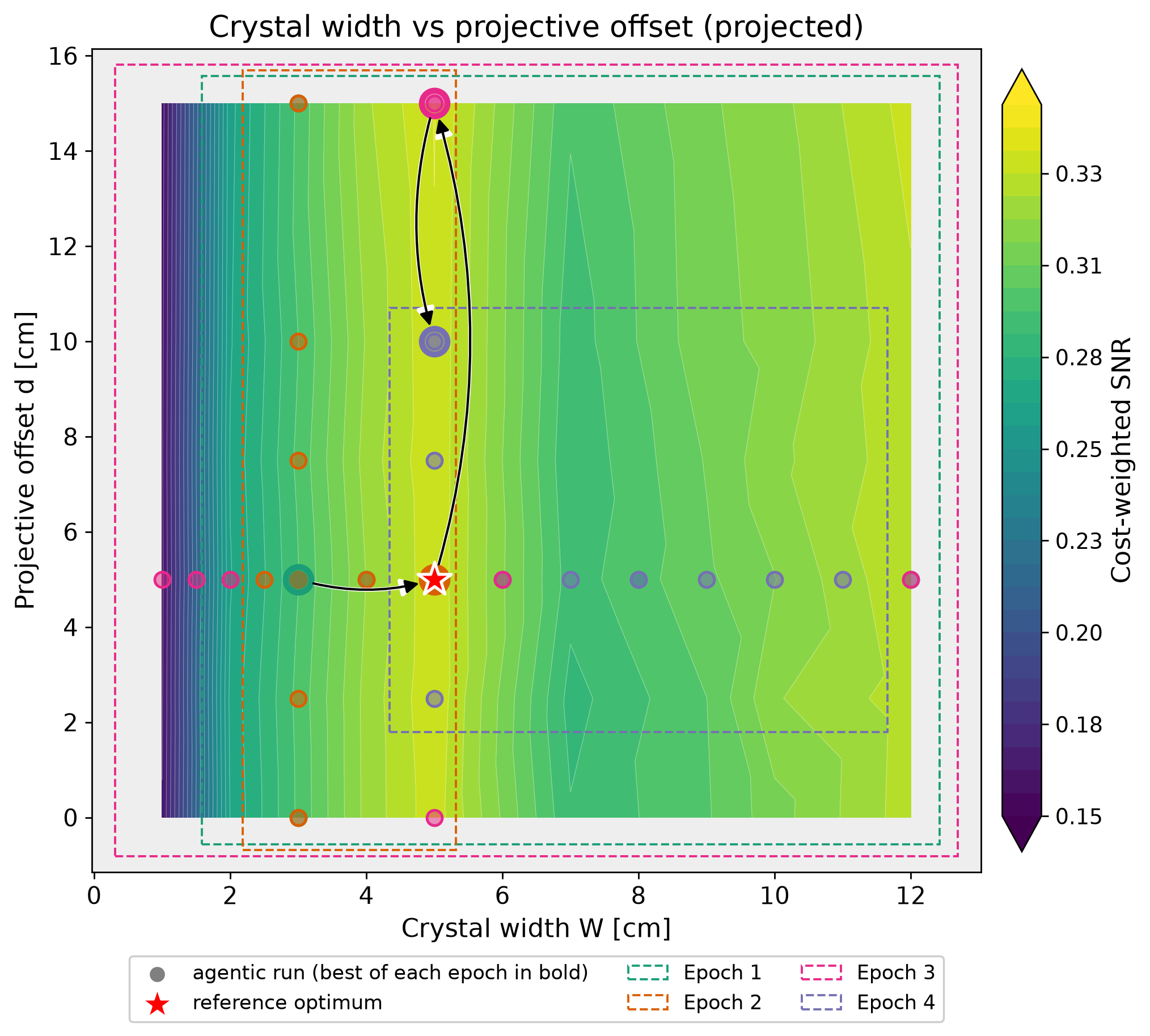}
    \includegraphics[width=0.52\textwidth]{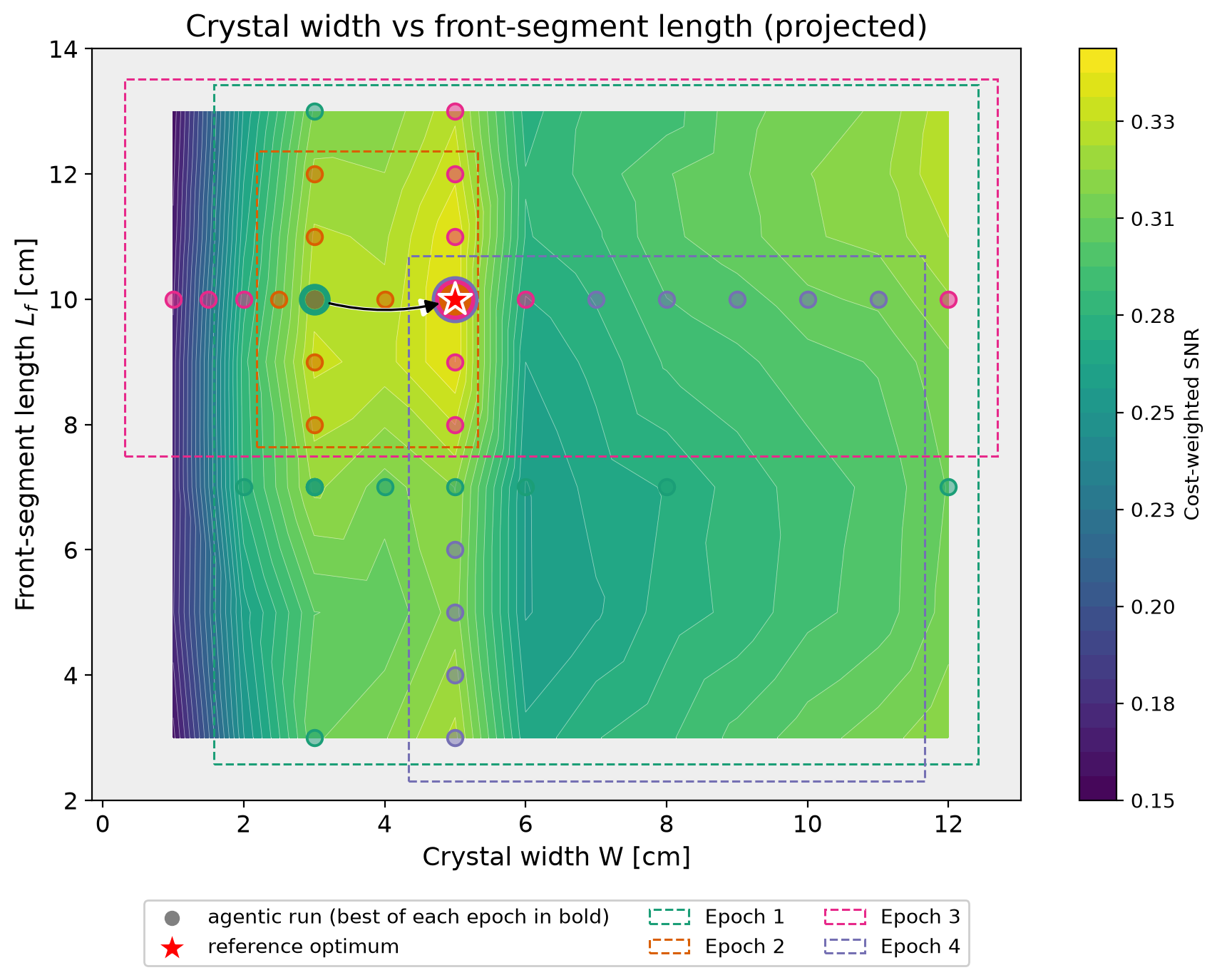}
    \includegraphics[width=0.5\textwidth]{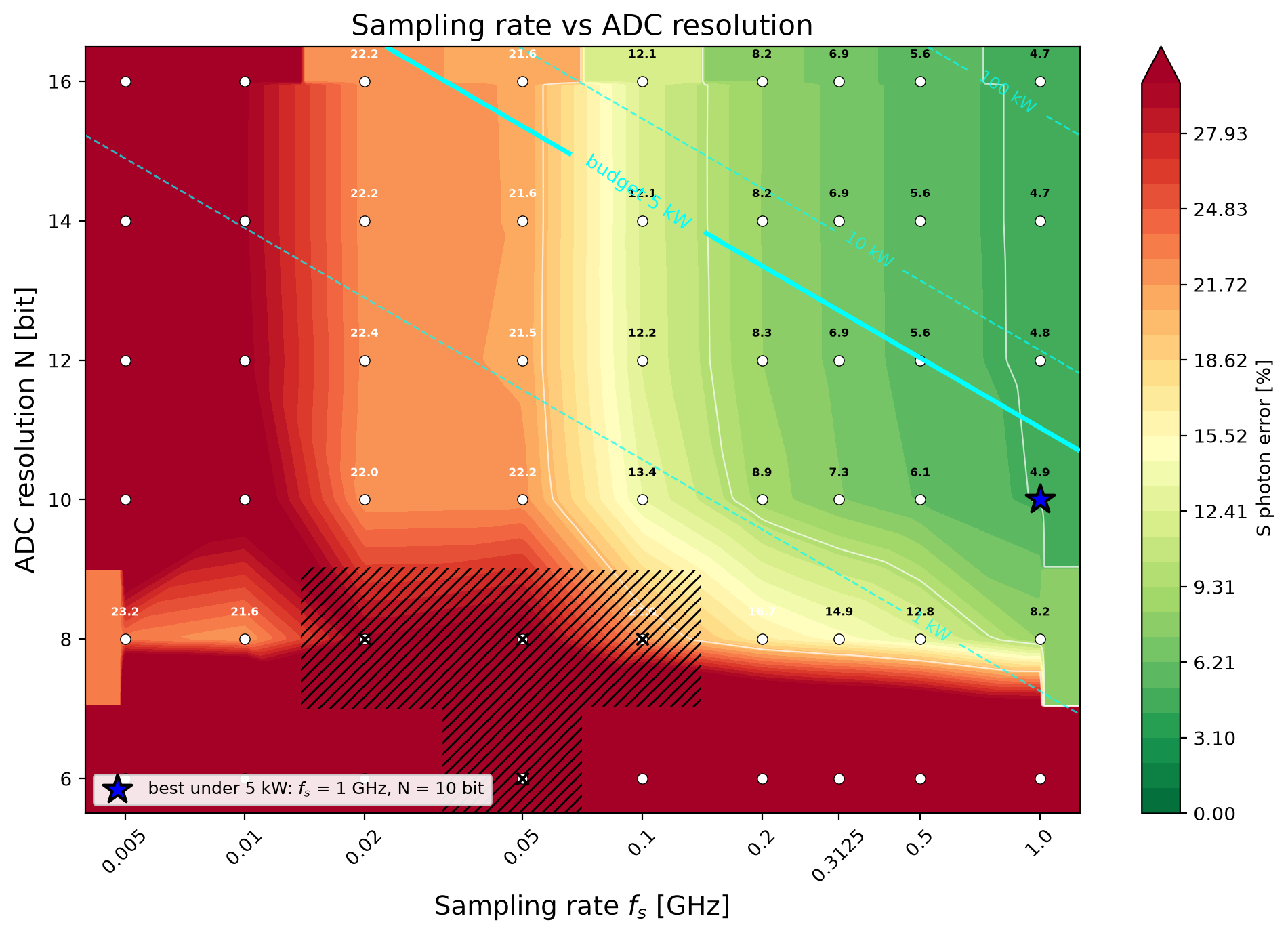}
    \caption{Agent-driven geometry optimization (one trajectory) and the subsequent digitization scan. (left) Crystal width versus projective offset and (center) crystal width versus front-segment length. The background is an interpolated reference slice of the cost-weighted SNR landscape from all simulated geometries (at $L_f = 12$ cm and $d = 5$ cm respectively); the markers are all geometries proposed by the agent, projected on the plane and coloured by epoch, so their colour is not the background value at that position. The best geometry of each epoch is drawn bold, the bold marker of the last epoch is the reported endpoint, dashed boxes are the coordinate extent of each epoch and arrows the epoch-to-epoch progression; the star is the reference point of the simulated set, $(W, L_f, d) = (5, 10, 5)$ cm, which is the true optimal in the given quantized parameter grid. The tested offsets between 5 and 15 cm are unresolved at given final width and length. (right) Sampling rate versus ADC resolution at the endpoint geometry: colour is the per-crystal error on the number of S photons,
    cyan lines the static ADC power, the solid line the 5 kW budget. The waveform template fitting fails at hatched cells. The star is the lowest error under the budget.}
    \label{fig:results_offset}
\end{figure}

\clearpage


\begin{ack}
This work is supported by the U.S. Department of Energy under contract numbers DE-AC02-76SF00515, DE-SC0007968, DE-SC0025956, and used resources of the National Energy Research Scientific Computing Center (NERSC), a Department of Energy User Facility using NERSC award HEP-ERCAP 0037461.
\end{ack}


\bibliographystyle{unsrt} 
\bibliography{main.bib}

@misc{butler2023report,
      title={{Report of the 2021 U.S. Community Study on the Future of Particle Physics (Snowmass 2021) Summary Chapter}}, 
      author={Joel N. Butler and R. Sekhar Chivukula and André de Gouvêa and Tao Han and Young-Kee Kim and Priscilla Cushman and Glennys R. Farrar and Yury G. Kolomensky and Sergei Nagaitsev and Nicolás Yunes and Stephen Gourlay and Tor Raubenheimer and Vladimir Shiltsev and Kétévi A. Assamagan and Breese Quinn and V. Daniel Elvira and Steven Gottlieb and Benjamin Nachman and Aaron S. Chou and Marcelle Soares-Santos and Tim M. P. Tait and Meenakshi Narain and Laura Reina and Alessandro Tricoli and Phillip S. Barbeau and Petra Merkel and Jinlong Zhang and Patrick Huber and Kate Scholberg and Elizabeth Worcester and Marina Artuso and Robert H. Bernstein and Alexey A. Petrov and Nathaniel Craig and Csaba Csáki and Aida X. El-Khadra and Laura Baudis and Jeter Hall and Kevin T. Lesko and John L. Orrell and Julia Gonski and Fernanda Psihas and Sara M. Simon},
      year={2023},
      eprint={2301.06581},
      archivePrefix={arXiv},
      primaryClass={hep-ex}
}

@article{Frank_2014,
doi = {10.1088/1742-6596/513/2/022010},
url = {https://doi.org/10.1088/1742-6596/513/2/022010},
year = {2014},
month = {jun},
publisher = {},
volume = {513},
number = {2},
pages = {022010},
author = {Frank, M and Gaede, F and Grefe, C and Mato, P},
title = {{DD4hep: A Detector Description Toolkit for High Energy Physics Experiments}},
journal = {Journal of Physics: Conference Series}
}

@inproceedings{BrunRademakers1996ROOT,
  author       = {Brun, Rene and Rademakers, Fons},
  title        = {{ROOT} - An Object Oriented Data Analysis Framework},
  booktitle    = {Proceedings of the AIHENP'96 Workshop},
  address      = {Lausanne},
  month        = sep,
  year         = {1996},
  note         = {Published in Nucl. Instrum. Meth. A 389 (1997) 81--86},
}

@ARTICLE{2020SciPy-NMeth,
  author  = {Virtanen, Pauli and Gommers, Ralf and Oliphant, Travis E. and
            Haberland, Matt and Reddy, Tyler and Cournapeau, David and
            Burovski, Evgeni and Peterson, Pearu and Weckesser, Warren and
            Bright, Jonathan and {van der Walt}, St{\'e}fan J. and
            Brett, Matthew and Wilson, Joshua and Millman, K. Jarrod and
            Mayorov, Nikolay and Nelson, Andrew R. J. and Jones, Eric and
            Kern, Robert and Larson, Eric and Carey, C J and
            Polat, {\.I}lhan and Feng, Yu and Moore, Eric W. and
            {VanderPlas}, Jake and Laxalde, Denis and Perktold, Josef and
            Cimrman, Robert and Henriksen, Ian and Quintero, E. A. and
            Harris, Charles R. and Archibald, Anne M. and
            Ribeiro, Ant{\^o}nio H. and Pedregosa, Fabian and
            {van Mulbregt}, Paul and {SciPy 1.0 Contributors}},
  title   = {{{SciPy} 1.0: Fundamental Algorithms for Scientific
            Computing in Python}},
  journal = {Nature Methods},
  year    = {2020},
  volume  = {17},
  pages   = {261--272},
  adsurl  = {https://rdcu.be/b08Wh},
  doi     = {10.1038/s41592-019-0686-2},
}

@misc{chung2025synthetictrainingrepresentationbridging,
      title={Synthetic Training and Representation Bridging in Reconstruction Domains}, 
      author={Wonyong Chung},
      year={2025},
      eprint={2505.05664},
      archivePrefix={arXiv},
      primaryClass={hep-ph},
      url={https://arxiv.org/abs/2505.05664}, 
}

@misc{birk2026scientifichumanagentreproductionpipeline,
      title={A Scientific Human-Agent Reproduction Pipeline}, 
      author={Joschka Birk and Gregor Kasieczka and Siddharth Mishra-Sharma and Benjamin Nachman and Dennis Noll and Tanvi Wamorkar},
      year={2026},
      eprint={2604.18752},
      archivePrefix={arXiv},
      primaryClass={hep-ph},
      url={https://arxiv.org/abs/2604.18752}, 
}

@book{p5_2023,
   title={Exploring the Quantum Universe: Pathways to Innovation and Discovery in Particle Physics},
   url={http://dx.doi.org/10.2172/2368847},
   DOI={10.2172/2368847},
   institution={Office of Scientific and Technical Information (OSTI)},
   author={Murayama, Hitoshi and Asai, Shoji and Heeger, Karsten and Ballarino, Amalia and Bose, Tulika and Cranmer, Kyle and Cyr-Racine, Francis-Yan and Demers, Sarah and Geddes, Cameron and Gershtein, Yuri and Heinemann, Beate and Hewett, JoAnne and Huber, Patrick and Mahn, Kendall and Mandelbaum, Rachel and Maricic, Jelena and Merkel, Petra and Monahan, Christopher and Onyisi, Peter and Palmer, Mark and Raubenheimer, Tor and Sanchez, Mayly and Schnee, Richard and Seidel, Sally and Seo, Seon-Hee and Thaler, Jesse and Touramanis, Christos and Vieregg, Abigail and Weinstein, Amanda and Winslow, Lindley and Yu, Tien-Tien and Zwaska, Robert},
   year={2023},
   month=june }

@techreport{CERN-ESU-2025-002,
      title         = "{The European Strategy for Particle Physics: 2026 Update -
                       Recommendations by the European Strategy Group}",
      reportNumber  = "CERN-ESU-2025-002",
      address       = "Monte Verità/Ascona, Switzerland",
      year          = "2025",
      url           = "https://cds.cern.ch/record/2950671",
      doi           = "10.17181/CERN.423R.S20Z",
}

@article{Moreno:2026mqk,
    author = "Moreno, Eric A. and Bright-Thonney, Samuel and Novak, Andrzej and Garcia, Dolores and Harris, Philip",
    title = "{AI Agents Can Already Autonomously Perform Experimental High Energy Physics}",
    eprint = "2603.20179",
    archivePrefix = "arXiv",
    primaryClass = "hep-ex",
    month = "3",
    year = "2026"
}

@article{RevModPhys.90.025002,
  title = {Dual-readout calorimetry},
  author = {Lee, Sehwook and Livan, Michele and Wigmans, Richard},
  journal = {Rev. Mod. Phys.},
  volume = {90},
  issue = {2},
  pages = {025002},
  numpages = {40},
  year = {2018},
  month = {Apr},
  publisher = {American Physical Society},
  doi = {10.1103/RevModPhys.90.025002},
  url = {https://link.aps.org/doi/10.1103/RevModPhys.90.025002}
}

@inproceedings{hirosky2024dualreadoutcalorimetryhomogeneouscrystals,
  author       = {Hirosky, R. and Anderson, T. and Cummings, G. and Dubnowski, M. and Guinto-Brody, C. and Guo, Y. and Ledovskoy, A. and Levin, D. and Madrid, C. and Martin, C. and Zhu, J.},
  title        = {Dual-readout calorimetry with homogeneous crystals},
  booktitle    = {Proceedings of CALOR2024},
  series       = {EPJ Web of Conferences},
  year         = {2024},
  note         = {arXiv:2408.11973},
  url          = {https://arxiv.org/abs/2408.11973}
}

@article{Eno:2025ltc,
    author = "Eno, S. and Wu, L. and Aamir, M. Y. and Chekanov, S. V. and Nabili, S. and Palmer, C.",
    title = "{On the resolution of dual readout calorimeters}",
    eprint = "2501.15329",
    archivePrefix = "arXiv",
    primaryClass = "physics.ins-det",
    doi = "10.1016/j.nima.2025.171080",
    journal = "Nucl. Instrum. Meth. A",
    volume = "1083",
    pages = "171080",
    year = "2026"
}

@misc{albertsson2019machinelearninghighenergy,
      title={Machine Learning in High Energy Physics Community White Paper},  
      author={Kim Albertsson and others},
      year={2019},
      eprint={1807.02876},
      archivePrefix={arXiv},
      primaryClass={physics.comp-ph},
      url={https://arxiv.org/abs/1807.02876}, 
}

@misc{aarrestad2026buildingainativeresearchecosystem,
      title={Building an AI-native Research Ecosystem for Experimental Particle Physics: A Community Vision}, 
      author={Thea Klaeboe Aarrestad and others},
      year={2026},
      eprint={2602.17582},
      archivePrefix={arXiv},
      primaryClass={hep-ex},
      url={https://arxiv.org/abs/2602.17582}, 
}

@inproceedings{Yilmaz_2025_ML4PS,
  title     = {Edge Machine Learning for Cluster Counting in Next-Generation Drift Chambers},
  author    = {Yilmaz, D. and Wu, L. and Gonski, J. and Rankin, D. and Herwig, C.},
  booktitle = {Proceedings of the Machine Learning for the Physical Sciences Workshop at NeurIPS 2025},
  year      = {2025},
  note      = {arXiv:2511.10540},
  url       = {https://arxiv.org/abs/2511.10540}
}

@article{Yoo_2024,
doi = {10.1088/2632-2153/ad6a00},
url = {https://doi.org/10.1088/2632-2153/ad6a00},
year = {2024},
month = {aug},
publisher = {IOP Publishing},
volume = {5},
number = {3},
pages = {035047},
author = {Yoo, Jieun and Dickinson, Jennet and Swartz, Morris and Di Guglielmo, Giuseppe and Bean, Alice and Berry, Douglas and Blanco Valentin, Manuel and DiPetrillo, Karri and Fahim, Farah and Gray, Lindsey and Hirschauer, James and Kulkarni, Shruti R and Lipton, Ron and Maksimovic, Petar and Mills, Corrinne and Neubauer, Mark S and Parpillon, Benjamin and Pradhan, Gauri and Syal, Chinar and Tran, Nhan and Wen, Dahai and Young, Aaron},
title = {Smart pixel sensors: towards on-sensor filtering of pixel clusters with deep learning},
journal = {Machine Learning: Science and Technology}
}

@misc{mode_collaboration_website,
  author       = {{MODE Collaboration}},
  title        = {MODE: Machine-learning Optimized Design of Experiments},
  year         = {2026},
  howpublished = {\url{https://mode-collaboration.github.io/}},
}

@article{Aehle_2025,
   title={Progress in end-to-end optimization of fundamental physics experimental apparata with differentiable programming},
   volume={13},
   ISSN={2405-4283},
   url={http://dx.doi.org/10.1016/j.revip.2025.100120},
   DOI={10.1016/j.revip.2025.100120},
   journal={Reviews in Physics},
   publisher={Elsevier BV},
   author={Aehle, Max and Arsini, Lorenzo and Barreiro, R. Belén and Belias, Anastasios and Boldyrev, Alexey and Bury, Florian and Cebrian, Susana and Demin, Alexander and Dickinson, Jennet and Donini, Julien and Dorigo, Tommaso and Doro, Michele and Gauger, Nicolas R. and Giammanco, Andrea and Gray, Lindsey and González, Borja S. and Kain, Verena and Kieseler, Jan and Kusch, Lisa and Liwicki, Marcus and Maier, Gernot and Nardi, Federico and Ratnikov, Fedor and Roussel, Ryan and Austri, Roberto Ruiz de and Sandin, Fredrik and Schenk, Michael and Scarpa, Bruno and Silva, Pedro and Strong, Giles C. and Vischia, Pietro},
   year={2025},
   month=dec, pages={100120} }

@misc{qasim2024physicsinstrumentdesignreinforcement,
      title={Physics Instrument Design with Reinforcement Learning}, 
      author={Shah Rukh Qasim and Patrick Owen and Nicola Serra},
      year={2024},
      eprint={2412.10237},
      archivePrefix={arXiv},
      primaryClass={physics.ins-det},
      url={https://arxiv.org/abs/2412.10237}, 
}

@article{Baydin02012021,
author = {Atılım Güneş Baydin and Kyle Cranmer and Pablo de Castro Manzano and Christophe Delaere and Denis Derkach and Julien Donini and Tommaso Dorigo and Andrea Giammanco and Jan Kieseler and Lukas Layer and Gilles Louppe and Fedor Ratnikov and Giles C. Strong and Mia Tosi and Andrey Ustyuzhanin and Pietro Vischia and Hevjin Yarar},
title = {Toward Machine Learning Optimization of Experimental Design},
journal = {Nuclear Physics News},
volume = {31},
number = {1},
pages = {25--28},
year = {2021},
publisher = {Taylor \& Francis},
doi = {10.1080/10619127.2021.1881364},
URL = {https://doi.org/10.1080/10619127.2021.1881364},
eprint = { https://doi.org/10.1080/10619127.2021.1881364}
}

@misc{schwartz2026resummationcparametersudakovshoulder,
      title={Resummation of the C-Parameter Sudakov Shoulder Using Effective Field Theory}, 
      author={Matthew D. Schwartz},
      year={2026},
      eprint={2601.02484},
      archivePrefix={arXiv},
      primaryClass={hep-ph},
      url={https://arxiv.org/abs/2601.02484}, 
}

@article{AGOSTINELLI2003250,
title = {Geant4—a simulation toolkit},
journal = {Nuclear Instruments and Methods in Physics Research Section A: Accelerators, Spectrometers, Detectors and Associated Equipment},
volume = {506},
number = {3},
pages = {250-303},
year = {2003},
issn = {0168-9002},
doi = {https://doi.org/10.1016/S0168-9002(03)01368-8},
url = {https://www.sciencedirect.com/science/article/pii/S0168900203013688},
author = {S. Agostinelli and J. Allison and K. Amako and J. Apostolakis and H. Araujo and P. Arce and M. Asai and D. Axen and S. Banerjee and G. Barrand and F. Behner and L. Bellagamba and J. Boudreau and L. Broglia and A. Brunengo and H. Burkhardt and S. Chauvie and J. Chuma and R. Chytracek and G. Cooperman and G. Cosmo and P. Degtyarenko and A. Dell'Acqua and G. Depaola and D. Dietrich and R. Enami and A. Feliciello and C. Ferguson and H. Fesefeldt and G. Folger and F. Foppiano and A. Forti and S. Garelli and S. Giani and R. Giannitrapani and D. Gibin and J.J. {Gómez Cadenas} and I. González and G. {Gracia Abril} and G. Greeniaus and W. Greiner and V. Grichine and A. Grossheim and S. Guatelli and P. Gumplinger and R. Hamatsu and K. Hashimoto and H. Hasui and A. Heikkinen and A. Howard and V. Ivanchenko and A. Johnson and F.W. Jones and J. Kallenbach and N. Kanaya and M. Kawabata and Y. Kawabata and M. Kawaguti and S. Kelner and P. Kent and A. Kimura and T. Kodama and R. Kokoulin and M. Kossov and H. Kurashige and E. Lamanna and T. Lampén and V. Lara and V. Lefebure and F. Lei and M. Liendl and W. Lockman and F. Longo and S. Magni and M. Maire and E. Medernach and K. Minamimoto and P. {Mora de Freitas} and Y. Morita and K. Murakami and M. Nagamatu and R. Nartallo and P. Nieminen and T. Nishimura and K. Ohtsubo and M. Okamura and S. O'Neale and Y. Oohata and K. Paech and J. Perl and A. Pfeiffer and M.G. Pia and F. Ranjard and A. Rybin and S. Sadilov and E. {Di Salvo} and G. Santin and T. Sasaki and N. Savvas and Y. Sawada and S. Scherer and S. Sei and V. Sirotenko and D. Smith and N. Starkov and H. Stoecker and J. Sulkimo and M. Takahata and S. Tanaka and E. Tcherniaev and E. {Safai Tehrani} and M. Tropeano and P. Truscott and H. Uno and L. Urban and P. Urban and M. Verderi and A. Walkden and W. Wander and H. Weber and J.P. Wellisch and T. Wenaus and D.C. Williams and D. Wright and T. Yamada and H. Yoshida and D. Zschiesche}
}

@misc{chung_bilevel_det_opt,
  author       = {Chung, Wonyong},
  title        = {{bilevel\_det\_opt}: Bilevel Optimization of Detector Geometry and Reconstruction Algorithm Parameters},
  year         = {2026},
  publisher    = {GitHub},
  journal      = {GitHub repository},
  howpublished = {\url{https://github.com/wonyongc/bilevel_det_opt}},
  note         = {Accessed: 2026-04-16}
}

@article{Lucchini:2020bac,
    author = "Lucchini, Marco T. and Chung, Wonyong and Eno, Sarah C. and Lai, Yihui and Lucchini, Lorenzo and Nguyen, Minh-Thi and Tully, Christopher G.",
    title = "{New perspectives on segmented crystal calorimeters for future colliders}",
    eprint = "2008.00338",
    archivePrefix = "arXiv",
    primaryClass = "physics.ins-det",
    doi = "10.1088/1748-0221/15/11/P11005",
    journal = "JINST",
    volume = "15",
    number = "11",
    pages = "P11005",
    year = "2020"
}

@misc{liu2026scifisafelightweightuserfriendly,
      title={SciFi: A Safe, Lightweight, User-Friendly, and Fully Autonomous Agentic AI Workflow for Scientific Applications}, 
      author={Qibin Liu and Julia Gonski},
      year={2026},
      eprint={2604.13180},
      archivePrefix={arXiv},
      primaryClass={cs.AI},
      url={https://arxiv.org/abs/2604.13180}, 
}

\appendix
\section{Replica trajectories}
\label{app:replicas}

All trajectories use the same prompt template; the agent is a fresh invocation per epoch that receives only the template, its own earlier axis declarations and the table of its own results. 
The reported endpoint is the best geometry of the last proposal epoch; the agents' own final statements applied an additional cost tie-break and are not used here.

\begin{figure}[!htbp]
    \centering
    \includegraphics[width=0.48\textwidth]{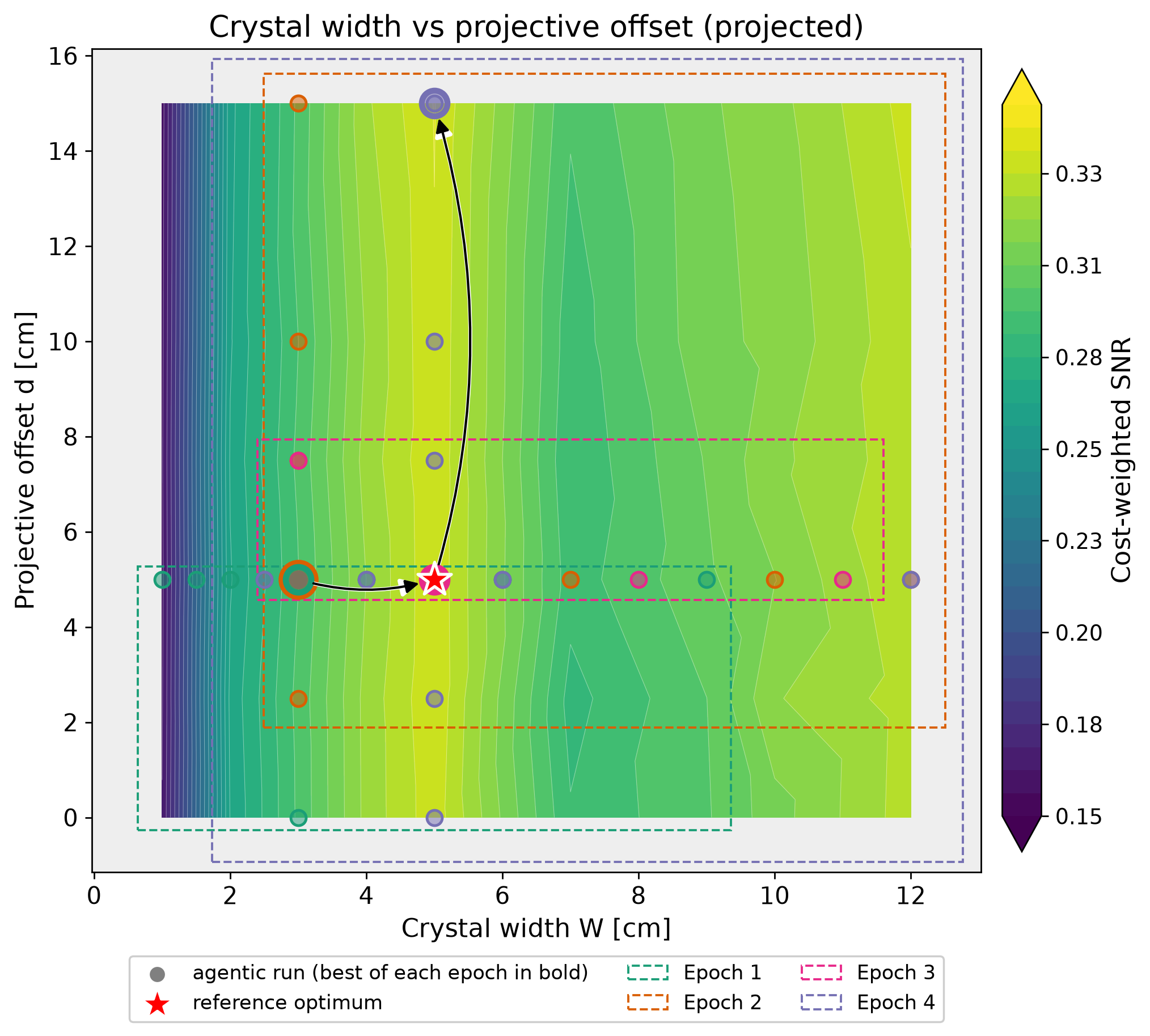}
    \includegraphics[width=0.48\textwidth]{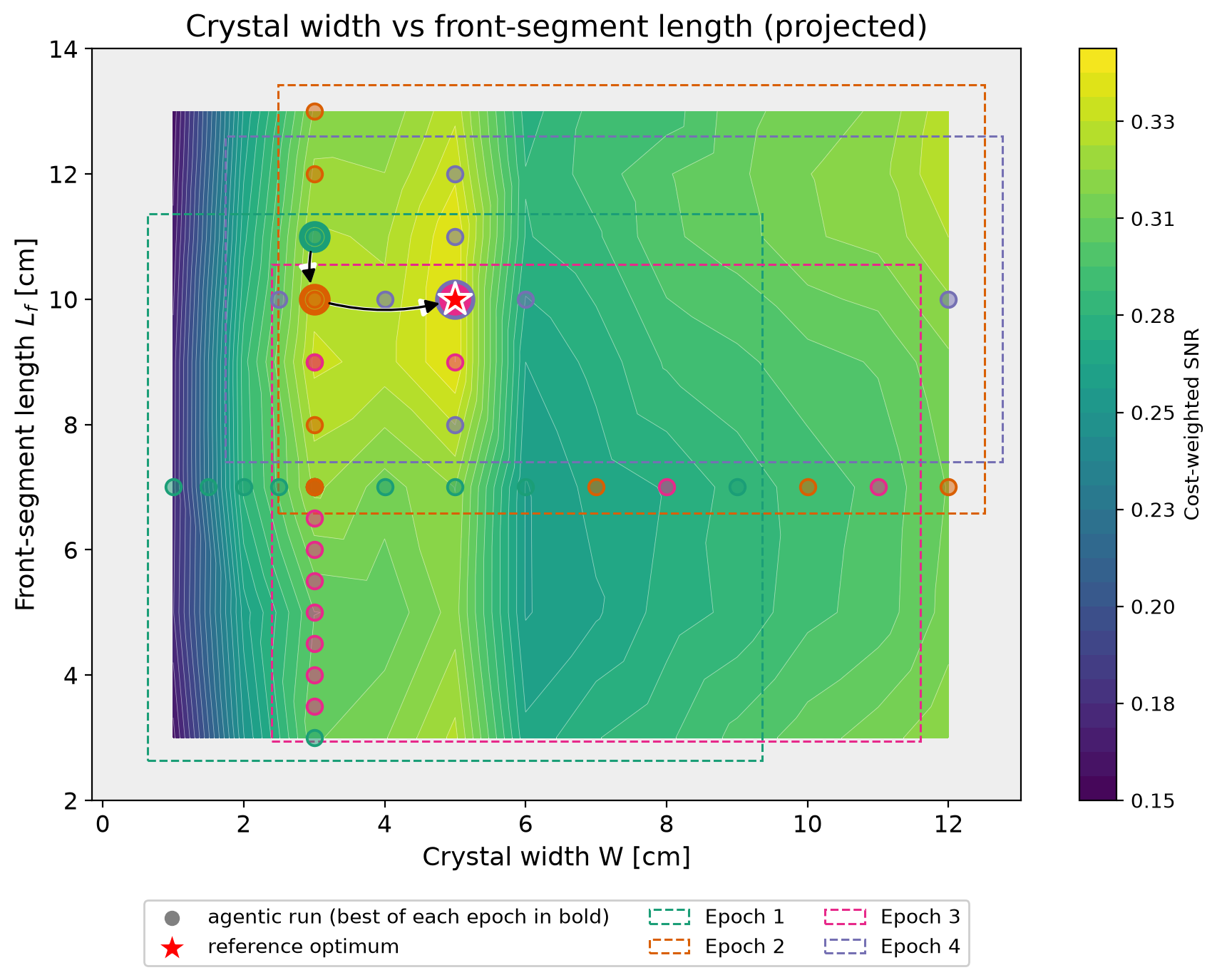}\\
    \includegraphics[width=0.48\textwidth]{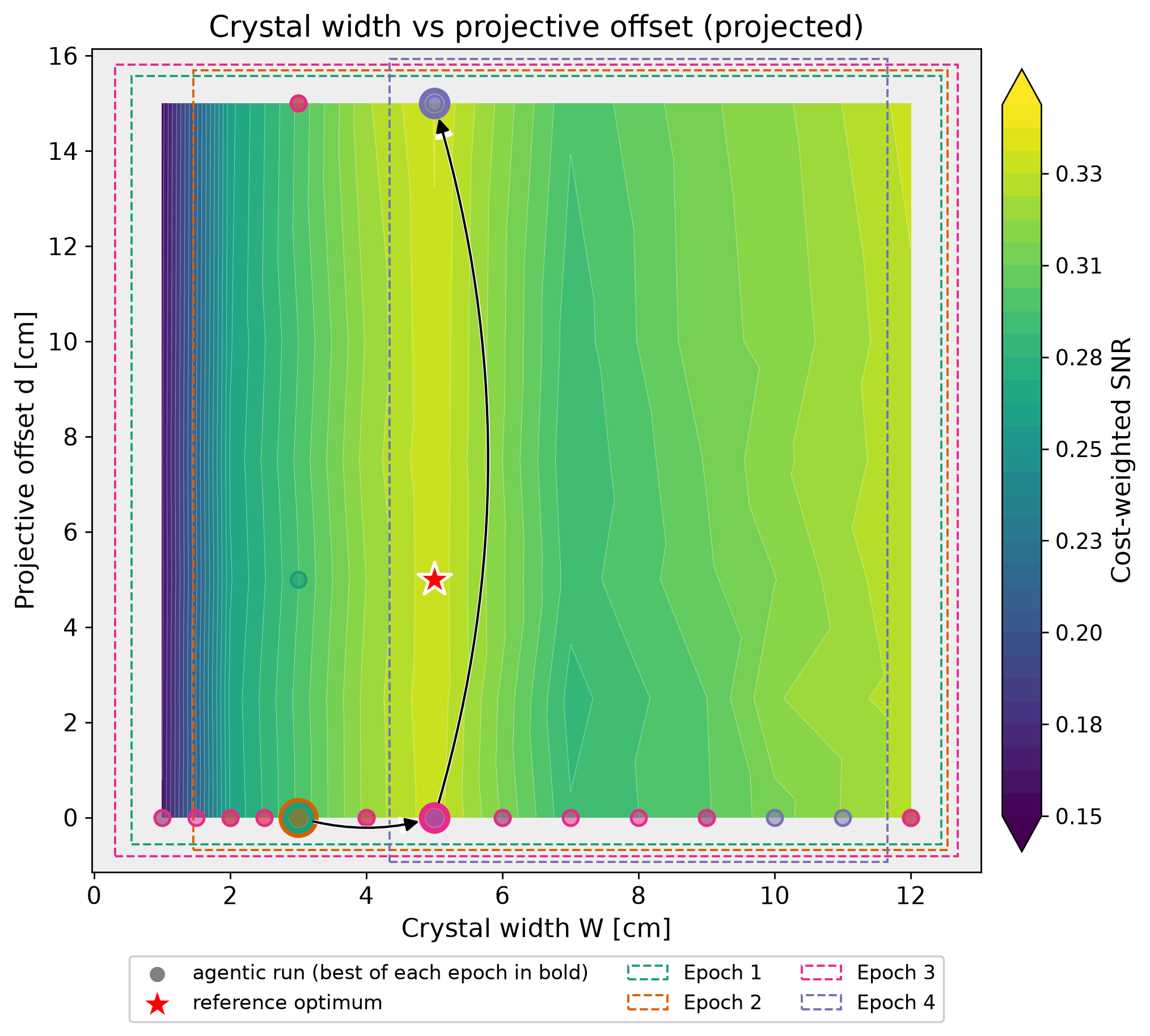}
    \includegraphics[width=0.48\textwidth]{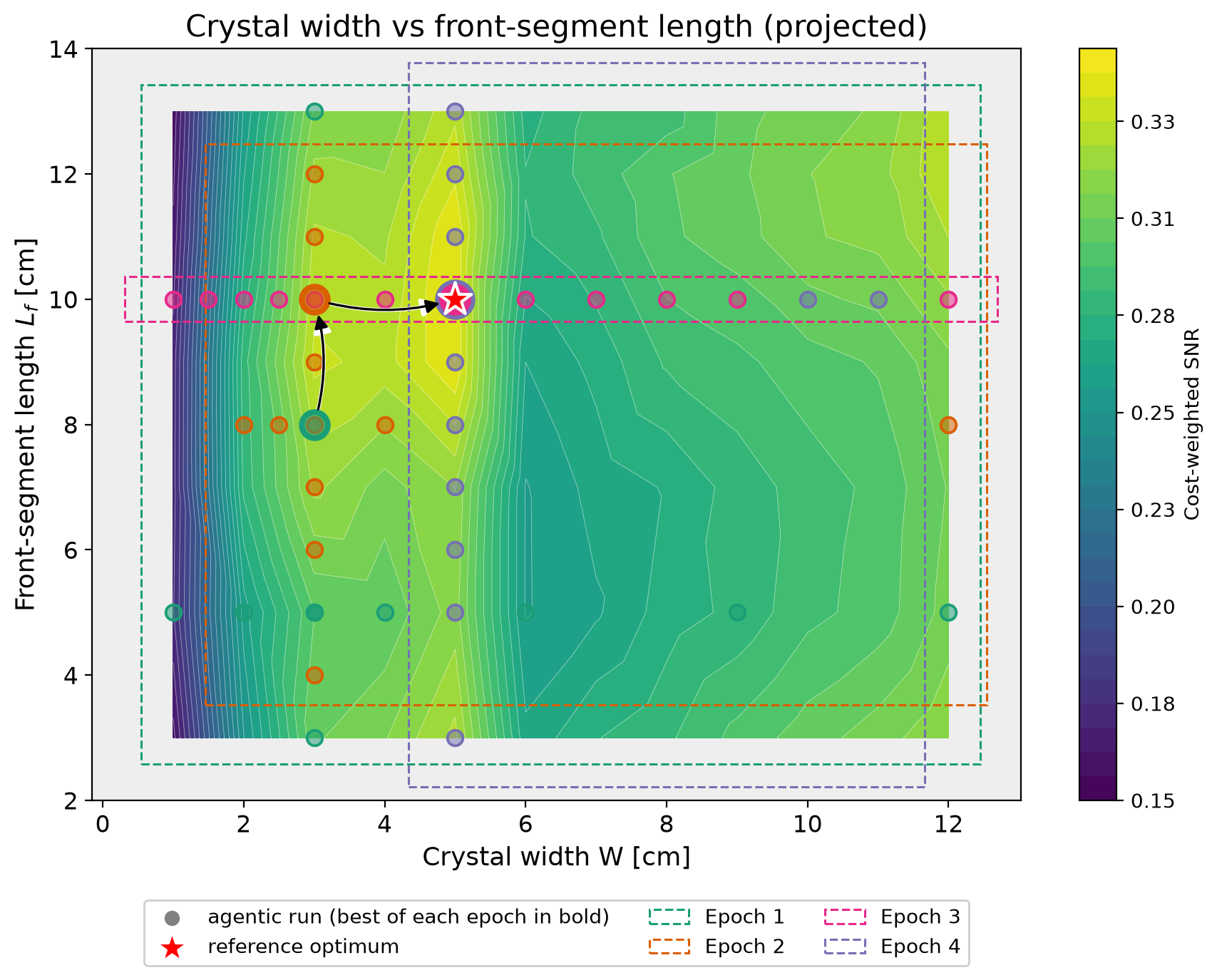}
    \caption{The two other trajectories of the five-epoch protocol (conventions as in Figure~\ref{fig:results_offset}). Both end at $W = 5$ cm, $L_f = 10$ cm, $d = 15$ cm; trajectory 1 (top) started at offset 5 cm and trajectory 2 (bottom) at offset 0 cm.}
    \label{fig:replicas_long}
\end{figure}




\end{document}